\documentclass[aps,showpacs,prl,twocolumn]{revtex4}
\usepackage{amssymb,amsfonts,amsmath}
\usepackage[usenames]{xcolor}
\usepackage{graphicx}
\usepackage{microtype}
\usepackage{epsfig}
\usepackage{amsmath}
\usepackage{setspace}
\usepackage{layout}
\usepackage{textcomp}
\usepackage{hyperref}
\usepackage{amsfonts}

\begin{document}
\bibliographystyle{apsrev}

\title{Critical values in Bak--Sneppen type models}
\author{Daniel Fraiman}
\affiliation{Departamento de Matem\'atica y Ciencias, Universidad de San Andr\'es, Buenos Aires, Argentina,}
\affiliation{CONICET, Argentina.}
\email{dfraiman@udesa.edu.ar}

\begin{abstract}
In the Bak--Sneppen model, the lowest fitness particle and its two nearest neighbors are renewed at each temporal step with a uniform (0,1) fitness distribution. The model presents a critical value that depends on the interaction criteria (two nearest neighbors) and on the update procedure (uniform). Here we calculate the critical value for models where one or both properties are changed. We study models with non-uniform updates, models with random neighbors and models with binary fitness and obtain exact results for the average fitness and for $p_c$.
\end{abstract}
\pacs{05.65.+b, 87.10.-e, 87.23-n, 89.75-k, 02.50.Cw}
\maketitle

The Bak Sneppen model~\cite{bak93} consists of $N$ particles located in a one-dimensional ring. Each particle $k$ 
is characterized by a quantity $X_k$, called fitness, that evolves in the following way:
\begin{equation}\label{BS}
X_k(t+1)=\left\{
\begin{array}{lll}
X_k(t) &  &   \mbox{if}\ \ dist(k,\tilde{k}_t)>a \\
U_{k,t} &  & \mbox{if}\ \   dist(k,\tilde{k}_t) \leq a. 
\end{array}
\right.
\end{equation}  
where $\tilde{k}_t=\{k:  \ \  X_k(t) \leq X_j(t)  \ \ \forall j \in \{1,2,\dots, N\}\}$ is the site with the lowest $X$ value at time $t$, 
 $U_{k,t}$ are iid random variables with uniform distribution in (0,1), and  $a \in \mathbb{N}$ is the number of neighbors on each side that are interacting with any given particle. The initial condition is uniform ($X_k(0)=U_{k,0}$) for all particles. The model is studied in the stationary regime. For large $N$ and $t$ most of the particles have a fitness uniformly distributed in $(p_c,1)$ \cite{bak93,bak95,boer,grassberger,paczuski,boettcher,felici}, only a few particles, $Z$, are below $p_c$ and these are participating in an avalanche. At the thermodynamic limit, all particles have a fitness uniformly distributed in $(p_c,1)$ since the number of particles that participate in an avalanche stabilize as $N$ grows, i.e.  $\underset{N\to \infty}{lim} \langle Z \rangle=cte$. Recently~\cite{daniel} we have shown that the critical value $p_c$ verifies: 
 \begin{equation}\label{fund}
p_c=\frac{1+\langle S\rangle}{1+2a},\\
\end{equation}
where  $\langle S \rangle$ is the mean number of interacting neighbors that have a fitness value below $p_c$ when the lowest fitness particle is below $p_c$. The term on the left of this equation describes the proportion of particles below $p_c$ after the update, while the term on the right describes the same proportion before the update.  In this letter, we study the BS model under different modifications. 
First, we study the model under a non-uniform update distribution. Second, we study the case of random neighbors for the update process. Finally, we study the binary fitness model. We also advance the description of $p_c$ and the average fitness.

First, we discuss what happens when the uniform updates assumption is broken.  Let us replace $U_k$ in eq. 1 by $W_{k,t}$, where $W_{k,t}$ are now independent and identically distributed continuous random variables with arbitrary p.d.f. $f(w)>0$ for $w\in \mathbb{R}$. The initial uniform condition is also replaced by  $W_{k,0}$.  Let $F(w)=\int_0^wf(h)dh$ be the cumulative probability function. For facilitation purposes, we use $Y$ to denote the new fitness value, and the dynamics remain as before
\begin{equation}\label{BSNU}
Y_k(t+1)=\left\{
\begin{array}{lclclc|}
Y_k(t) &  &   \mbox{if}\ \ dist(k,\tilde{k}_t)>a \\
W_{k,t} &  & \mbox{if}\ \   dist(k,\tilde{k}_t) \leq a, 
\end{array}
\right.
\end{equation}  
 where now $\tilde{k}_t$ is the lowest fitness particle at time $t$. As expected, this model also exhibits self-organized criticality, since no major modifications are applied to the original model. In fact, it is easy to see that the joint probability of $(Y_1(t),Y_2(t),...,Y_N(t))$ is the same as that of  $(F^{-1}(X_1(t)),F^{-1}(X_2(t)),...,F^{-1}(X_N(t)))$ for all $t$. Therefore, at equilibrium ($t\to \infty$), it is enough to understand the uniform fitness case (eq.~\ref{BS}) to extrapolate to an arbitrary fitness distribution (eq.~\ref{BSNU}). In the first case, once the system reaches equilibrium, particle fitness $X$ is uniform$(p_c,1)$. Hence, at the thermodynamic limit, particles that evolve with non-uniform updates converge to a situation where the fitness is greater than a critical value $p^{nu}_c$ and it satisfies
\begin{equation}\label{fund_af}
p^{nu}_c=F^{-1}(p_c)=F^{-1}(\frac{1+\langle S\rangle}{1+2a}).
\end{equation}
Moreover, applying $F^{-1}$ one can see that the fitness of the particles converge to a p.d.f. $h$ equal to
\begin{equation}\label{fund_af}
h(y)=\left\{
\begin{array}{lll}
0 &  &   \mbox{if}\ \ y<p^{nu}_c\\
\\
f(y)/(1-p^{nu}_c)&  & \mbox{if}\ \  y \geq p^{nu}_c,
\end{array}
\right.
\end{equation}
just by applying $F^{-1}$ to the uniform case.  We emphasize that $\langle S\rangle$ does not depend on $f(w)$, it depends only on $a$.  This result implies that if one chooses a non-uniform update distribution that favors small values of fitness (e.g. $F(1/2)>1/2$) then the critical value $p^{nu}_c$ will be smaller than $p_c$. Fig. 1 highlights the difference between uniform and non-uniform update distributions.

   \begin{figure}
\begin{center}
\includegraphics[width=0.5\textwidth]{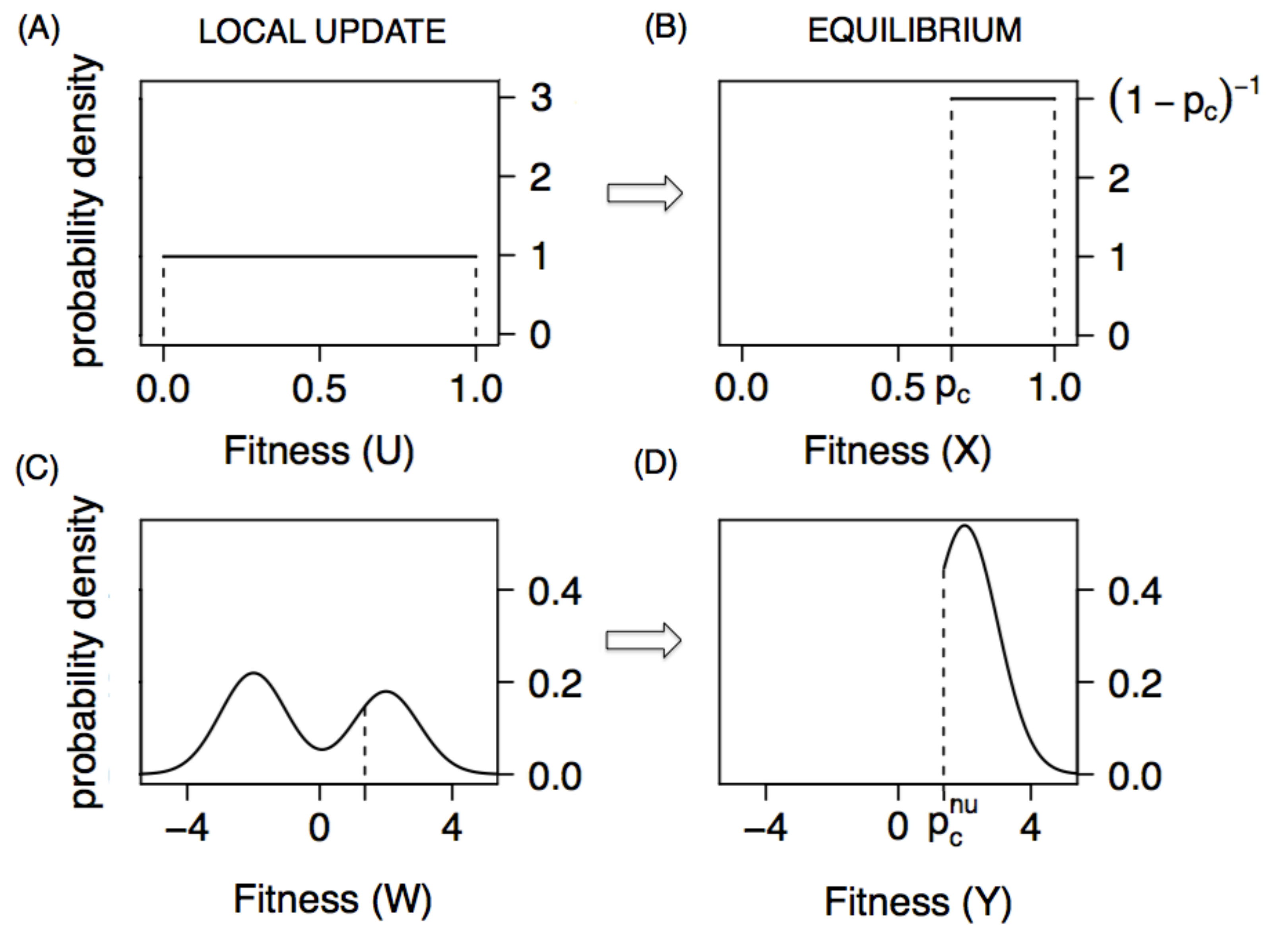}
\caption{Bak--Sneppen model. The update distribution is shown on the left and the equilibrium fitness distribution on the right. Two cases are shown: the (A) uniform (0,1) update distribution and (C) a two modes update distribution example, $f(x)$.  The dashed line in (C) corresponds to $p_c^{nu}$ , which verifies $\int^{p_c^{nu}}_{-\infty}f(x)dx=p_c$. }\label{fig1}
\end{center}
\end{figure}
Given the difficulty of obtaining exact results in the Bak--Sneppen (BS) model, we study it under different modifications that make it more  manageable. Next, we discuss some of these simplified models and study the impact of using equation~\ref{fund} and transformations similar to the one presented for the non-uniform update on those models. 

\paragraph{Model 1: Random interacting neighbors.}

\begin{figure}[h]
\begin{center}
\includegraphics[width=0.4\textwidth]{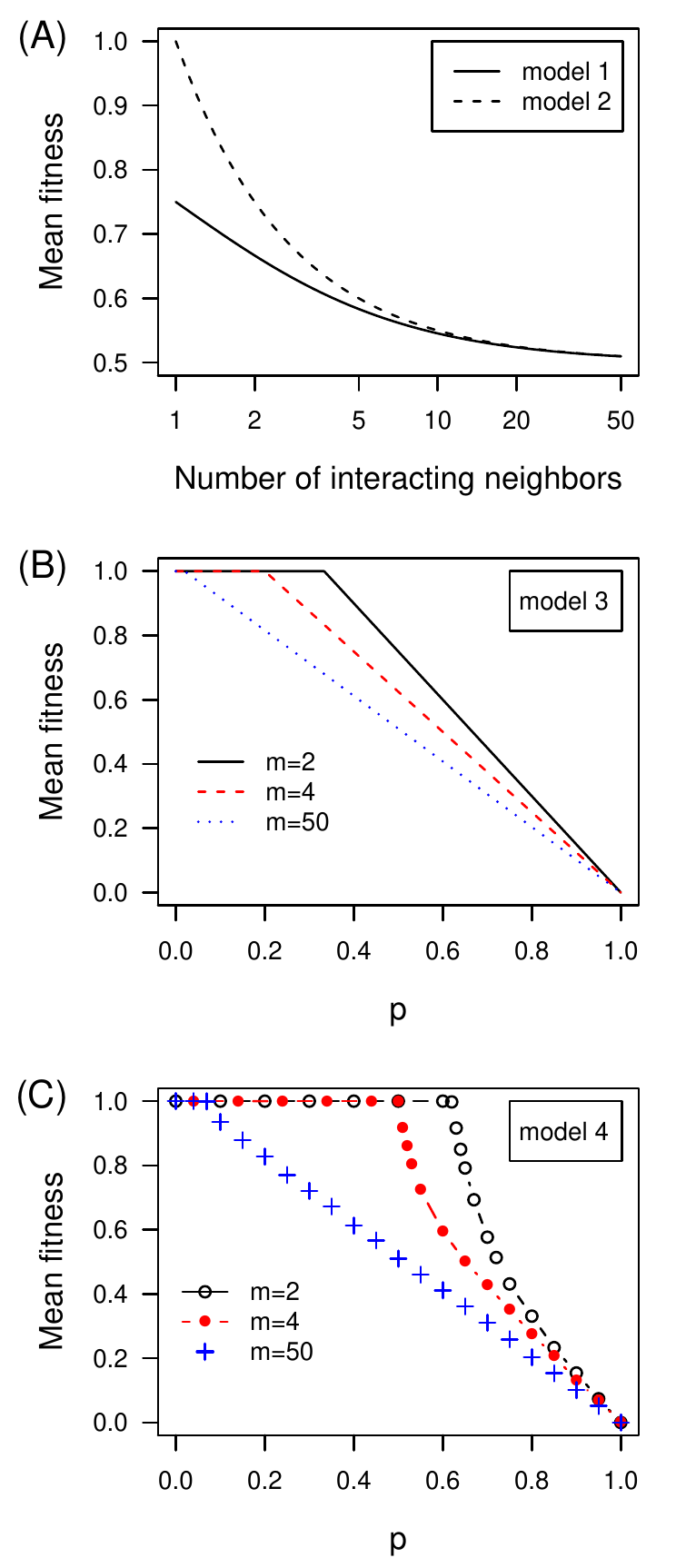}
\caption{ A) Mean fitness of models 1 and 2 as a function of the number of interacting neighbors. 
Mean fitness of (B) model 3 and (C) model 4 as a function of p (probability of fitness equal to zero immediately after the update).
}\label{fig:2cap6}
\end{center}
\end{figure}
The most well-known modification to the BS model is to break the assumption that the nearest neighbors are the ones that interact with the lowest fitness particle. A simplified hypothesis is that the interacting particles are chosen randomly between all possible particles at each time step. Let us call this model the random Bak--Sneppen model (rBS model).  This model also presents a critical value, $p^{r}_c$. In~\cite{meester} the authors found that the critical value is $p^{r}_c=1/2$ when considering only one interacting neighbor. 
 In~\cite{daniel}, we also studied this model and looked for a lower bound for $p_c$.  Ref. ~\cite{daniel} found that
\begin{equation}\label{fund2}
p^{r}_c=\frac{1}{1+2a}.
\end{equation}
just replacing $\langle S \rangle$ by zero in eq.~\ref{fund}. The value of $\langle S \rangle$ is zero because at the thermodynamic limit, all particles have a fitness value greater than $p_c$ except for the ones that are part of the avalanche, which are not each others' neighbors. Fitness converges to a uniform $(p^r_c,1)$ distribution, thus the mean fitness is equal to $\frac{1+p^{r}_c}{2}$. Fig. 2A shows the mean fitness as a function of the number of interacting neighbors. 

Note that the number of interacting neighbors can be odd in this model; one just needs to replace $2a$ by the number of interacting neighbors, $m$.  This result can be extended to the case of non-uniform fitness.  The new critical value  $p^{r,nu}_c$ verifies 
\begin{equation}\label{rand}
p^{r,nu}_c=F^{-1}(p^{r}_c)=F^{-1}(\frac{1}{1+m}).
\end{equation}

\paragraph{Model 2: Random interacting neighbors and binary fitness}

A discrete fitness version of this model can be introduced just by considering values of fitness, now called $Y$, that can only take the values 0 or 1.  The dynamics is the following: at each discrete time, a random site with fitness value 0 is updated, as are $m$ other random sites. If there is no site with 0 fitness, a random site with a fitness of 1 is selected and updated with other $m$ random sites.  The sites are always updated with a Bernoulli variable with probability 1/2. Note that, as there may be ties in fitness, we are obligated to randomly select the lowest fitness particle.   The goal is the same as before: we want to understand the limit probability law of the particles' fitness. Since fitness in this case is binary, we study the proportion of particles with fitness equal 1, $P(Y=1)$, which is equivalent to $\langle Y \rangle$.

This model can be described in terms of the rBS model. Let us take the original rBS model with fitness values $X$ and apply the following function over the fitness 
\begin{equation}\label{psi}
\Psi_{1/2}(X)=\left\{
\begin{array}{lclc|cl}
1 &  &   \mbox{if}\ \  X > 1/2 \\
0 &  & \mbox{if}\ \    X  \leq 1/2. \\
\end{array}
\right.
\end{equation}
Again, the joint probability of $(Y_1(t),Y_2(t),...,Y_N(t))$ is the same as that of  $(\Psi_{1/2}(X_1(t)),\Psi_{1/2}(X_2(t)),...,\Psi_{1/2}(X_N(t)))$, i.e. 
The $\Psi_{1/2}$ function converts the rBS model into the discrete version introduced above. 
At the thermodynamic limit, once the system reaches equilibrium, we know that the fitness of the rBS model particles obeys a uniform distribution from $p^{r}_c$ to 1. Therefore, the proportion of particles that have a discrete fitness equal to 1, $\langle Y \rangle$,  verifies  
\begin{equation}\label{equ}
\underset{N\to \infty}{ \lim} \langle Y \rangle=\underset{N\to \infty}{ \lim} \langle \Psi_{1/2}(X) \rangle=P(X>1/2)=\frac{1+m}{2m}.\\
\end{equation} 
Note that for $m=1$ all particles have a discrete fitness equal to 1.  Fig. 2A shows the behavior of the mean fitness as a function of the number of interacting neighbors (eq.~\ref{equ}).

\paragraph{Model 3: Random interacting neighbors, binary fitness and Bernoulli updates}

A new model can be introduced if we consider that the updates of the discrete fitness values obey a Bernoulli variable with probability $1-p$ of having fitness equal to 1. This model, unlike the Bak--Sneppen model, has a parameter. It resembles a percolation model more than the \textit{self-organized} Bak--Sneppen model. Nevertheless, since part of the Bak-Sneppen dynamics is conserved (dynamics governed by the minimum fitness value) we anticipate some unexpected behavior at a particular value of $p$. 
Fortunately, as before, this model can be obtained from the rBS model. Just by applying the function $\Psi_{p}$ (interchange 1/2 by $p$ in eq.~\ref{psi}) to the particle fitness that evolves according to the rBS model, we can obtain the new binary fitness.  

At the thermodynamic limit, the fraction of sites with fitness equal to 1 behaves in the following way with the parameter $p$, 
\begin{equation*}
\underset{N\to \infty}{ \lim} \langle Y \rangle=\underset{N\to \infty}{ \lim}\langle \Psi_{p}(X ) \rangle=\left\{
\begin{array}{lclc|cl}
\frac{1-p}{1-p^{r}_c} &  &   \mbox{if}\ \  p > p^{r}_c \\
1 &  & \mbox{if}\ \    p \leq p^{r}_c.\\
\end{array}
\right.
\end{equation*} 
which is equivalent (using eq.~\ref{fund2}) to 
\begin{equation}
\underset{N\to \infty}{ \lim} \langle Y \rangle=\left\{
\begin{array}{lll}
(1+\frac{1}{m})(1-p)  &  &   \mbox{if}\ \  p > \frac{1}{1+m} \\
1 &  & \mbox{if}\ \    p \leq \frac{1}{1+m}.
\end{array}
\right.
\end{equation} 
Fig. 2B shows the average fitness as a function of $p$ for three different values of $m$.

\paragraph{Model 4: Nearest neighbor interactions, binary fitness and Bernoulli updates}

\begin{figure}
\begin{center}
\includegraphics[width=0.5\textwidth]{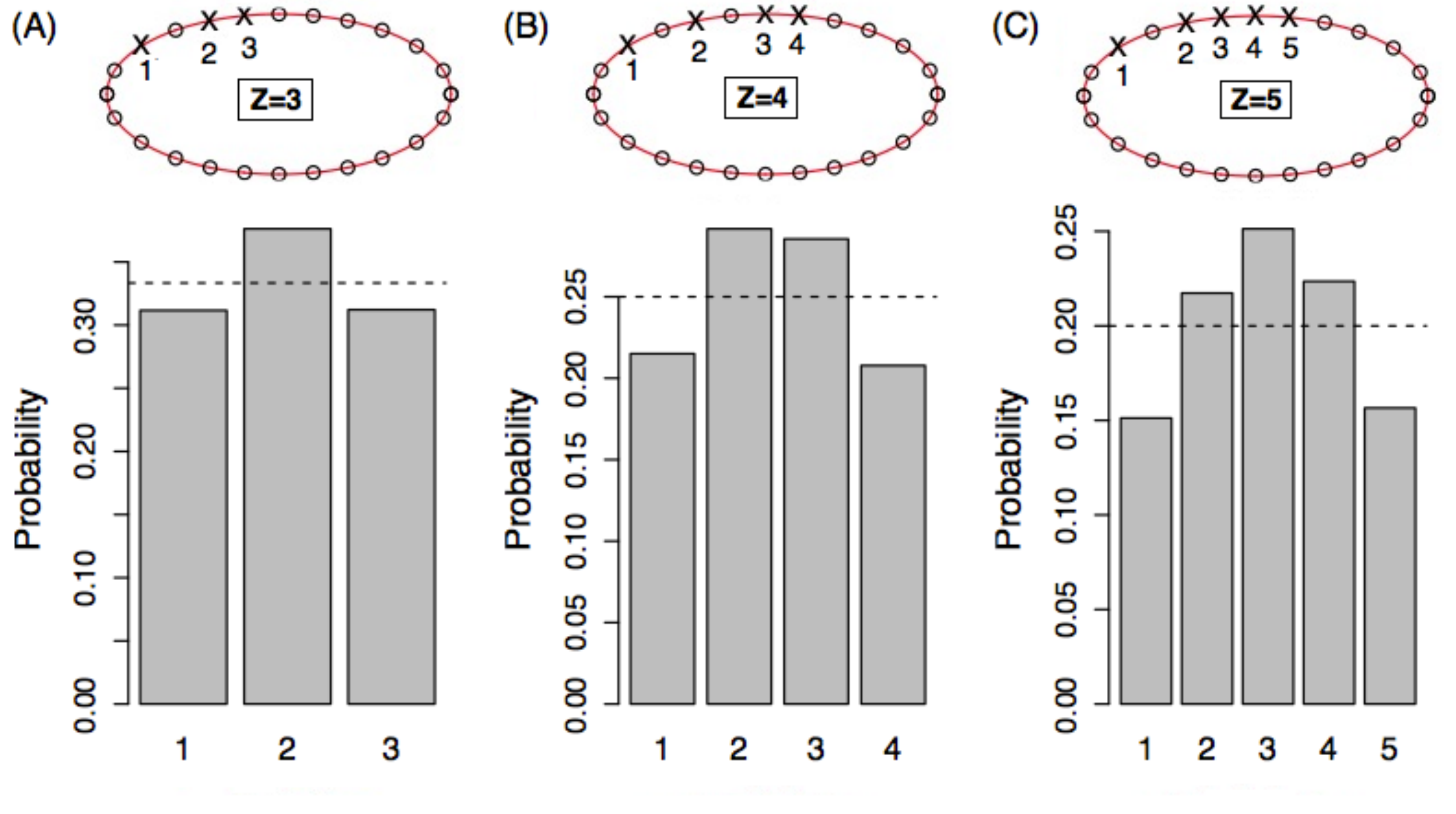}
\caption{ Simulations from the BS model for $m=2$. Upper panels: Representation of the BS model where circles represent inactive particles and crosses represent active ones. 
Lower panels: Proportion of times the lowest fitness particle is the one in the $i-th$ position going clockwise when there are  (A) three, (B) four or (C) five active particles.}\label{fig3}
\end{center}
\end{figure}

Finally, we discuss a model similar to Model 3 where the nearest neighbors are the ones that are updated. This model was introduced by Barbay and Kenyon~\cite{barbay}. Fitness values are binary, but now, once the lowest fitness particle is selected (randomly, since there are ties), the $m$ nearest neighbors are updated with independent Bernoulli variables with parameter $1-p$~\footnote{In~\cite{barbay} the authors updated with Bernoulli variables with parameter $p$ and studied the case $m=2$.}.  
In~\cite{barbay} the authors show that for $m=2$ a critical value $p^{BK}_c$ exists. Moreover, they prove~\cite{barbay} that $0.4563<p^{BK}_c$ and show by simulations that $p^{BK}_c\approx 0.635$. 

Here, we present a better lower bound as well as an upper bound for the case of an arbitrary number of neighbors ($m$). The result is the following,   
\begin{equation}\label{cotas}
p^{r}_c< p^{BK}_c<p_c.
\end{equation}
The critical value of the BK model is bounded by the critical values of the rBS and BS models. For $m=2$ eq.~\ref{cotas} state that $1/2< p^{BK}_c<2/3$. The lower bound $p^{r}_c< p^{BK}_c$ is easy to understand based on the results presented in~\cite{daniel}. We have already shown that more ``compact'' avalanches give rise to larger critical values. Now, if we compare the avalanches of Model 3 (critical value $p^{r}_c$) with the ones generated with Model 4 (critical value $p^{BK}_c$) we see that the latter are more compact, since Model 4 evolves through the updates of \textit{nearest} neighbors. To understand the upper bound, $p^{BK}_c<p_c$, the argument is more complex but is based on the same idea as avalanche compaction. The avalanches of the discrete version of the BS model are more compact than the ones of the BK model. We explain this argument next.

We first describe the BK model mathematically. Let $N_{0,t}$ be the set that contains all particles with fitness equal to zero at time $t$, and $N_{1,t}$ the set that contains the rest of the particles (with fitness equal to 1). Let us randomly choose one particle in each set (with equal probability), which we call $h_{0,t}$, and $h_{1,t}$ to the randomly selected particle from set $N_{0,t}$ and $N_{1,t}$, respectively. The lowest fitness particle at time $t$, $\tilde{k}_t$, is defined as  
 \begin{equation}\label{mini}
\tilde{k}_t=\left\{
\begin{array}{lclc|cl}
h_{0,t} &  &   \mbox{if} \ \ N_{0,t}\neq  \varnothing    \\
h_{1,t} &  &   \mbox{if} \ \  N_{0,t}= \varnothing.     \\
\end{array}
\right.
\end{equation}  
Particles obey the following dynamics 
 \begin{equation}\label{BK}
Y_k(t+1)=\left\{
\begin{array}{lll}
Y_k(t) &  &   \mbox{if}\ \ dist(k,\tilde{k}_t)>a \\
W_{k,t} &  & \mbox{if}\ \   dist(k,\tilde{k}_t) \leq a, 
\end{array}
\right.
\end{equation}  
where $W_{k,t}$ and $Y_k(0)$ are iid Bernoulli random variables with parameter $1-p$. For $p=1/2$, although the model seems to be the binary fitness version of the Bak-Sneppen model, it is not. In order to create its discrete version, there must exist a function $\Psi:\mathbb{R}\to \{0,1\}$ (or $\Psi:\mathbb{R}^N\to \{0,1\}^N$)  which, when applied individually to each fitness particle (or to its vector), converts the fitness given by eq.~\ref{BS} into the discrete version given by eq.~\ref{BK}. Specifically, if the BK model with $p=1/2$ were the discrete version of the BS model, then the joint probability of $(Y_1(t),Y_2(t),\dots,Y_N(t))$ must be equal to the joint probability of $(\Psi(X_1(t)),\Psi(X_2(t)),\dots,\Psi(X_N(t)))$ (or $\Psi(X_1(t),X_2(t),\dots,X_N(t)))$), where $X_k$ is the fitness of particle $k$ given by the BS model. To understand why there is no $\Psi$ function that verifies the previous conditions, we focus on the selection mechanism of the lowest fitness particle. In the rBS (Model 1), in its discrete version (Model 2), and also in Model 3, each particle below $p^r_c$ has the same probability of being the lowest one (the same happens above $p^r_c$). The BS model behaves in a different way. Each particle below $p_c$ does not have the same chance of being the lowest one. At equilibrium, knowing that there are $Z_t$ particles below $p_c$ at time $t$, the ones at the edge of the avalanche have a lower chance than $\frac{1}{Z_t}$ of being the lowest one, which is one of the reasons why avalanches in the BS model diffuse so slowly. Fig. 3 shows empirical evidence of this non-equiprobable law. Once the system is in equilibrium, we take snapshots and study the distribution of the fitness of active particles. In these snapshots, there are different numbers of active particles ($Z$). We only analyze the fitness of the snapshots that verify $Z=\{3,4,5\}$. Once we have the fitness values for $Z=3$, for example, we construct a vector $(X_a,X_b,X_c)$ with those fitness values. In the vector's first position, $X_a$, we put the fitness of the first clockwise particle; in the second position, we put the fitness of the particle in the middle, and in the third position, we put the last particle's fitness. The same procedure is done for $Z$ equals 4 and 5. Fig 3 shows the fraction of times each particle was the one with the lowest fitness when $Z=3$ (panel A), $Z=4$ (panel B) and $Z=5$ (panel C).
As one can observe, the distribution is symmetric as it must be, but the particles in the middle have a greater probability of being the lowest fitness particle. Moreover, the distance between the active particles shapes this distribution (data not shown), i.e. the probability of being the lowest particle depends on the relative positions of the active particles (I.e., the order and distance between them). In the BK model, all active particles have the same chance of being the lowest fitness particle, and in the BS model, we just showed that the particles in the middle have the greatest probability. This last observation is the key argument for why there is no $\Psi$ function that converts the BS model into the BK model preserving the dynamics. There is no way how to convert the non-equiprobable selection mechanism in a equiprobable one (eq. 12).

Nevertheless, if we apply the $\Psi_{p}$ function previously defined to the fitness of the BS model, we obtain the true discrete version of the BS model when $p=1/2$. This discrete model evolves by eq.~\ref{BK} but with $\tilde{k}_t$, obeying a much more complicated probability law than the one defined in eq.\ref{mini}. In particular $h_{0,t}$ is not selected with equal probability from the $N_{0,t}$ set.  The specific law for $\tilde{k}_t$ exceeds the scope of this work, but we can say that this true discrete version of the BS model will verify the same type of behavior found in Model 3,
\begin{equation*}
\underset{N\to \infty}{ \lim} \langle Y \rangle=\underset{N\to \infty}{ \lim}\langle \Psi_{p}(X ) \rangle=\left\{
\begin{array}{lll}
\frac{1-p}{1-p_c} &  &   \mbox{if}\ \  p > p_c \\
1 &  & \mbox{if}\ \    p \leq p_c.
\end{array}
\right.
\end{equation*}

On the other hand, it is difficult for the BK model (equations~\ref{mini} and~\ref{BK}) to calculate the exact behavior of $\underset{N\to \infty}{ \lim} \langle Y \rangle$ as a function of $p$. We can only say that $p^{BK}_c$ must be smaller than the critical value of the true discrete BS model, which is equal to the continuous case ($p_c$), since the true discrete BS model generates more compact avalanches than the BK model, and more compact avalanches give rise to larger critical values. 

Finally, we study $\underset{N\to \infty}{ \lim} \langle Y \rangle$ by simulations.   Fig 2 (C) shows the average fitness as a function $p$ for cases $m=\{2,4,50\}$. We can say that  
\begin{equation}
\underset{N\to \infty}{ \lim} \langle Y \rangle=\left\{
\begin{array}{lll}
g(p) &  &   \mbox{if}\ \  p > p^{BK}_c \\
1 &  & \mbox{if}\ \    p \leq p^{BK}_c,\\
\end{array}
\right.
\end{equation} 
where $g$ is a non-linear decreasing function. We believe that there is an interesting and challenging problem in describing function $g$ in detail.

In summary, in this paper we show how transformations of the original BS model can be done without altering the model's complex dynamics. First we showed how to compute the $p_c$ value for the BS model with non-uniform updates. Although perhaps not surprising, this first result gave us the mathematical tools to analyze four different Bak--Sneppen type models in detail. We studied the average fitness and critical values for all cases (see Fig. 2).

\end{document}